\documentclass{article}

\usepackage{graphicx}
\usepackage[strings]{underscore}
\usepackage{bm}
\usepackage{amsmath}
\usepackage{amsfonts}
\usepackage{mathtools}
\usepackage{subfig}
\usepackage{hyperref}
\hypersetup{
	colorlinks = true,
	linkcolor = blue,
	anchorcolor = blue,
	citecolor = blue,
	filecolor = blue,
	urlcolor = blue
}
\usepackage{cleveref}
\usepackage{epstopdf, epsfig}
\usepackage{color}
\usepackage[margin=1in]{geometry}

\newcommand*\diff{\mathop{}\!\mathrm{d}}

\title{Modelling inelastic granular media using Dynamical Density Functional Theory
}

\author{
	B. D. Goddard
		\footnote{\url{b.goddard@ed.ac.uk},School of Mathematics and the Maxwell Institute for Mathematical Sciences, University of Edinburgh,
		Edinburgh, UK, EH9 3FD} 
	\and
	T. D. Hurst
		\footnote{\url{t.hurst@sms.ed.ac.uk},School of Mathematics and the Maxwell Institute for Mathematical Sciences, University of Edinburgh,
		Edinburgh, UK, EH9 3FD} 
	\and
	R. Ocone
		\footnote{\url{r.ocone@hw.ac.uk}, School of Engineering and Physical Sciences, Heriot-Watt University, Edinburgh, UK, EH14 4AS}
}

\begin{document}
	\maketitle
\begin{abstract}
	We construct a new mesoscopic model for granular media using Dynamical Density Functional Theory (DDFT). The model includes both a collision operator to incorporate inelasticity and the Helmholtz free energy functional to account for external potentials, interparticle interactions and volume exclusion. We use statistical data from event-driven microscopic simulations to determine the parameters not given analytically by the closure relations used to derive the DDFT. We numerically demonstrate the crucial effects of each term in the DDFT, and the importance of including an accurately parametrised pair correlation function.
\end{abstract}
\section{Introduction}\label{Introduction}
Granular media is ubiquitous in industrial and natural processes \cite{Bagnold2005,Richard2005}, but very difficult to accurately model in large systems. Microscopic models \cite{Cundall1979,Bannerman2011} generally produce accurate results, but are usually restricted to simple systems, a relatively small number of particles, or short simulation times. This is due to computational cost scaling poorly with the number of particles in the system. Models which approximate the media as a macroscopic continuum \cite{Lun1984,vanWachem2003} are not inhibited by the number of particles in the system. However, continuum models need to be supplied with constitutive equations for bulk properties, such as the particle stress tensor. Such constitutive equations, in the dilute flow regime, are usually obtained by invoking the kinetic-collisional theory \cite{Lun1984}. The disadvantages of such constitutive equations lie mainly in what is believed to be its inability to treat the meso-scale: issues such as cluster formation (and breakage), for instance, have been discussed at large \cite{Louge2014}, and attempts to solve those issues have been proposed (e.g., by  “adjusting” the Navier-Stokes equations \cite{Louge2014,Mitrano2014}).

Recently, approaches using Dynamical Density Functional Theory (DDFT) have produced promising results in the field of modelling complex fluids \cite{Marconi2000,Archer2009,Lutsko2010} at a {\em mesoscopic} level. Derived from a particle-based model, DDFTs are continuum models that utilise the well-studied Helmholtz free energy functional \cite{Henderson1992} and can include interparticle and external potentials, volume exclusion \cite{Roth2010}, hydrodynamic interactions \cite{Rex2009,Goddard2012} and multiple species \cite{Goddard2013}. However, current models do not account for inelastic (or indeed elastic) collisional dynamics, which have crucial dissipative effects in granular media.

We introduce a DDFT adapted for granular media in \Cref{Section:Derivation} which incorporates particle collisions at the mesoscopic level, and present some numerical results in \Cref{Section:Numerics}, which show the potential of DDFT for modelling granular media. There are many areas for further investigation, which we discuss at the end of this paper.
\section{Derivation of the model}\label{Section:Derivation}
\subsection{Microscopic Dynamics}
A set of $N$ particles, each of mass $m$, in $d$ dimensions can be modelled via Langevin \cite{Langevin1905} or Newton equations
of motion: for positions $\bm{r}^N=(\bm{r}_1,\bm{r}_2,...,\bm{r}_N)$ and momenta $\bm{p}^N=(\bm{p}_1,\bm{p}_2,...,\bm{p}_N)$, the dynamics are given by
\begin{gather}
\frac{\diff \bm{r}^N}{\diff t}=\frac{\bm{p}^N}{m},\quad
\frac{\diff\bm{p}^N}{\diff t}=-\bm{\nabla}_{\bm{r}^N}V(\bm{r}^N,t)-\gamma\bm{p}^N+\bm{a}(t),\label{eq:Langevin}
\end{gather}
where
\begin{align}
V(\bm{r}^N,t) =  \sum_{i=1}^NV_{\text{ext}}(\bm{r}_i,t) 
+\frac{1}{2}\sum_{i \neq j=1}^NV_2(\bm{r}_i,\bm{r}_j,t)
+\frac{1}{6}\sum_{i \neq j \neq k=1}^NV_3(\bm{r}_i,\bm{r}_j,\bm{r}_k,t)
+\cdots .
\end{align}
Here $V_{\text{ext}}(\bm{r}_i,t)$ represents any external potential, for example gravity. Pairwise interactions are modelled by an interparticle potential $V_2(\bm{r}_i,\bm{r}_j,t)$ for $i,j=1,...,N$, and we analogously include higher order interparticle potentials such as $V_3(\bm{r}_i,\bm{r}_j,\bm{r}_k)$ for $i,j,k=1,...,N$. The second term on the right hand side of the equation for momentum in \cref{eq:Langevin} represents external frictional effects, for example if the particles are moving in a bath, where $\gamma>0$ determines the strength of the effect. The final term $\bm{a}(t)$ is a Brownian motion term \cite{Brown1828} from thermal fluctuations in the bath, with strength $(mk_BT\gamma)^{1/2}$ determined by a fluctuation-dissipation theorem \cite{Ermak1978}, where $k_B$ is Boltzmann's constant and $T$ is the bath temperature. Granular media particles are usually assumed to be unaffected by thermal fluctuations, however for modelling purposes this term is sometimes included as a thermostat \cite{Marini2007}. 

For point-like particles, \cref{eq:Langevin} fully describes the dynamics of the system. In this paper, we assume that particles are spherical with diameter $\sigma$. We include the effects of collisions in the dynamics of particles by restricting their movement to the {\em hard sphere domain}:
\begin{align}
\mathcal{D}_N=\{\bm{r}^N,\bm{p}^N\in\mathbb{R}^{dN}:\forall i,j, \|\bm{r}_i-\bm{r}_j\|\ge\sigma\}.\label{eq:HardSphereDomain}
\end{align}
When two particles that are moving toward one another come into contact, we must then instantaneously change their momenta to avoid particle overlap. We assume that collisions are binary and instantaneous; {\em i.e.}\ a collision between the $i^{\mathrm{th}}$ and $j^{\mathrm{th}}$ particles occurs at time $t$ if $\|\bm{r}_i(t)-\bm{r}_j(t)\|=\sigma$, and $(\bm{p}_i(t)-\bm{p}_j(t))\cdot(\bm{r}_i(t)-\bm{r}_j(t))<0$. To resolve the collision we apply a collision rule. A standard collision rule which maps pre-collisional velocities $\bm{p}_i^{\mathrm{in}},\bm{p}_j^{\mathrm{in}}$ post-collisional velocities $\bm{p}_i^{\mathrm{out}},\bm{p}_j^{\mathrm{out}}$ is given by \cite{Bannerman2009}:
\begin{align}
\bm{p}_i^{\text{out}}=\bm{p}_{i}^{\text{in}}-\frac{(1+\alpha)}{2}\bm{\nu}^{i,j}\cdot(\bm{p}_i^{\text{in}}-\bm{p}_j^{\text{in}})\bm{\nu}^{i,j},\quad
\bm{p}_j^{\text{out}}=\bm{p}_{j}^{\text{in}}+\frac{(1+\alpha)}{2}\bm{\nu}^{i,j}\cdot(\bm{p}_i^{\text{in}}-\bm{p}_j^{\text{in}})\bm{\nu}^{i,j},\label{eq:CollisionRule}
\end{align}
where $\bm{\nu}^{i,j}=(\bm{r}_i-\bm{r}_j)/\|\bm{r}_i-\bm{r}_j\|$ and $\alpha\in(0,1]$ is the coefficient of restitution. If $\alpha=1$, the collisions are perfectly elastic and no energy is dissipated, and the component of velocity in the direction of the collision is reflected. If $\alpha<1$, energy is lost via a reduction of the velocity component in the direction of the collision. We note that linear and angular momentum are conserved by this collision rule:
\begin{gather}
\bm{p}_i^{\mathrm{out}} + \bm{p}_j^{\mathrm{out}} 
=
\bm{p}_i^{\mathrm{in}} + \bm{p}_j^{\mathrm{in}},\\
(\bm{r}_i-\bm{x})\times\bm{p}_i^{\mathrm{out}} 
+(\bm{r}_j-\bm{x})\times\bm{p}_j^{\mathrm{out}} 
=
(\bm{r}_i-\bm{x})\times\bm{p}_i^{\mathrm{in}} 
+(\bm{r}_j-\bm{x})\times\bm{p}_j^{\mathrm{in}} ,\quad\forall\bm{x}\in\mathbb{R}^3.
\end{gather}
However, kinetic energy is not conserved when $\alpha<1$:
\begin{gather}
\|\bm{p}_i^{\mathrm{out}}\|^2 + \|\bm{p}_j^{\mathrm{out}}\|^2 \le \|\bm{p}_i^{\mathrm{in}}\|^2 + \|\bm{p}_j^{\mathrm{in}}\|^2 
\end{gather}
For simplicity, in this derivation we do not investigate the effect of angular momentum on the DDFT model \cite{Duran2016} when collisional effects are included.

In principle, \cref{eq:Langevin}, on the domain \cref{eq:HardSphereDomain} with collision rule \cref{eq:CollisionRule} can accurately model a system of $N$ particles. Particle based methods can produce very accurate results, but for large $N$, or when the system is very dense, reaching the desired simulation time is generally infeasible, or the simulation becomes too memory-intensive. Furthermore, when $\alpha<1$ a system of particles obeying the above microscopic dynamics can experience inelastic collapse, where an infinite number of collisions can occur in finite time, effectively jamming simulations \cite{McNamara1994}.

\subsection{$N$ body equations for rigid particles}
If collisions are neglected, associated with \cref{eq:Langevin} is the {\em Kramer's equation} (or in absence of thermal fluctuations, the {\em Liouville equation}) \cite{Risken1996}, a partial differential equation (PDE) which models the dynamics of the $N$-particle distribution function $f^{(N)}(\bm{r}^N,\bm{p}^N,t)$, the probability of finding $N$ particles with positions $\bm{r}^N$ and momenta $\bm{p}^N$ at time $t$:
\begin{align}
\Bigg[\frac{\partial}{\partial t}+\frac{1}{m}\bm{p}\cdot\bm{\nabla}_{\bm{r}^N}&-\bm{\nabla}_{\bm{r}^N}V(\bm{r}^N,t)\cdot\bm{\nabla}_{\bm{p}^N}\Bigg]f^{(N)}(\bm{r}^N,\bm{p}^N,t)\nonumber\\
&-\bm{\nabla}_{\bm{p}^N}\cdot\left[\gamma(\bm{p}^N+mk_BT\bm{\nabla}_{\bm{p}^N})f^{(N)}(\bm{r}^N,\bm{p}^N,t)\right]=0.\label{eq:Kramers}
\end{align}
When constructing \cref{eq:Kramers} for deterministic dynamics, the microscopic dynamics are assumed to be smooth. However, particles which undergo instantaneous collisions have discontinuities in their velocity profile, and so their trajectories are not smooth. In \cite{Wilkinson2018}, the weak formulation of the Liouville equation is derived for a system of elastic spherical particles that obey linear dynamics, using distribution theory. Without any additional assumptions, careful dissection of the phase space in the weak formulation leads to an additional collisional term:
\begin{align}
C[f^{(N)}] = \int_{\partial\mathcal{D}_N(\bm{r}_1,t)}\int_{\mathbb{R}^{dN}}f^{(N)}(\bm{r}^N,\bm{p}^N,t)\bm{p}^N\cdot\hat{\bm{n}}\diff\bm{p}^N\diff \mathcal{H}(\bm{r}),\label{eq:KramersCollisionTerm}
\end{align}
where for $k\le N$,
\begin{align}
\mathcal{D}_N(\bm{r}^k,t)&=\{\bm{r}^{N-k}\in\mathbb{R}^{d(N-k)},\bm{p}^N\in\mathbb{R}^{dN}:\forall i,j, \|\bm{r}_i-\bm{r}_j\|\ge\sigma\},\label{eq:ReducedHardSphereDomain}
\end{align}
and $\hat{\bm{n}}$ is the outward unit normal of $\partial\mathcal{D}_N(\bm{r}_1,t)$ with $\mathcal{H}(\bm{r}^N)$ the Hausdorff measure on $\partial\mathcal{D}_N(\bm{r}_1,t)$. When considering the associated Bogoliubov-Born-Green-Kirkwood-Yvon (BBGKY) hierarchy \cite{Balescu1975} (in weak form), the additional term integrates to the well-known Boltzmann collision operator for $\alpha=1$.  We note that this is an alternative method to derivations where the collision term is constructed by additional assumptions on an interaction force at the level of the BBGKY hierarchy \cite{Huang1987}.

It is a topic of future work to see how long range potentials, friction and inelasticity affect this derivation. In this work we make an assumption that is popular in the literature; interactions between particles can be split into a term which describes `soft' interactions, and a term for `hard' interactions, {\em e.g.} a collision operator. This assumption is heuristically validated by the derivation in \cite{Wilkinson2018} where the collision operator occurs due to geometric properties of the phase space. The collision operator derived in \cite{Wilkinson2018} does not include soft interparticle interactions, so the potential term $V_2$ must be incorporated separately.

\Cref{eq:Kramers} (or an analogous weak formulation including \cref{eq:KramersCollisionTerm}) replaces a system of $2N$ differential equations with a single PDE. However, the (spatial-momentum) dimension of $f^{N}(\bm{r}^N,\bm{p}^N,t)$ is $2dN$. If we were to simulate this system on a discretized domain with $M$ points in each direction, we would require $M^{2dN}$ points for simulation, which quickly becomes computationally intractable. However, it is known \cite{Chan2005} that the $N$-particle distribution function is a functional of the one-body position density:
\begin{align}
\rho(\bm{r}_1,t) = N\int_{\mathbb{R}^{(N-1)d}}
\int_{\mathbb{R}^{Nd}}
\diff\bm{p}^N\diff\bm{r}^{N-1}f^{(N)}(\bm{r}^N,\bm{p}^N,t)\chi_{\mathcal{D}_N(\bm{r}^1,t)}.\label{eq:density}
\end{align}
We include the characteristic function to stress that the phase space of the system does not allow particles to overlap, ensuring that we integrate over the hard-sphere domain while keeping the first position variable free. In contrast, in derivations where particles interact purely via soft potentials, the integral in position is over $\mathbb{R}^{d(N-1)}$. Heuristically, when integrating \cref{eq:Kramers} the inclusion of this characteristic function leads to collisional terms in the BBGKY hierarchy. Rigorously, the weak formulation admits the Boltzmann collision operator \cite{Wilkinson2018}.

To arrive at an equation to model $\rho(\bm{r},t)$ we first define the {\em $n$-reduced phase space particle distribution function} by
\begin{align}
f^{(n)}(\bm{r}^n,\bm{p}^n,t)=\frac{N!}{(N-n)!}\int_{\mathbb{R}^{(N-n)d}}\int_{\mathbb{R}^{(N-n)d}} \diff\bm{r}^{(N-n)} \diff\bm{p}^{(N-n)}f^{(N)}(\bm{r}^N,\bm{p}^N,t)\chi_{\mathcal{D}_N(\bm{r}^{n},t)},\label{eq:ReducedDistributions}
\end{align}
where $\bm{r}^{N-n} = (\bm{r}_{n+1},...,\bm{r}_N)$, $\bm{r}^{n}=(\bm{r}_1,\bm{r}_2,...,\bm{r}_{n})$, and similar for $\bm{p}^{N-n},\bm{p}^{n}$. To ease notation in the derivation we write $\bm{r}=\bm{r}_1$, $\bm{p}=\bm{p}_1$. By integrating \cref{eq:Kramers} with respect to $\bm{r}^{N-1}$ and $\bm{p}^{N-1}$, and appealing to the symmetry of arguments in $f^{(N)}$, we arrive at the {\em one-body Kramer's equation}, the first equation in the BBGKY hierarchy:
\begin{align}
\frac{\partial f^{(1)}}{\partial t}+\frac{1}{m}\bm{p}\bm{\cdot\nabla_{r}}f^{(1)}-&\bm{\nabla_{r}}V^{ext}(\bm{r},t)\bm{\cdot\nabla_\bm{p}}f^{(1)}
\nonumber-\bm{\nabla_{\bm{p}}\cdot}\left[\gamma(\bm{p}+mk_BT\nabla_{\bm{p}})f^{(1)}\right]\\
&-\frac{1}{m}\mathcal{L}_{\mathrm{coll}}(f^{(2)})
-\mathcal{L}_{\mathrm{part}}(f^{(2)},f^{(3)},...,f^{(N)})
=0,\label{eq:OneBodyKramers}
\end{align}
where $\mathcal{L}_{\text{coll}}(f^{(2)})$ incorporates binary collisions via a collision operator. We consider the following inelastic collision operator \cite{vanBeijeren1983}:
\begin{align}
\mathcal{L}_{\mathrm{coll}}(f^{(2)}) = &\nonumber
\sigma^{d-1}
\int_{\mathbb{R}^d}\diff\bm{p}_2
\int_{\mathbb{S}^{d-1}}
\diff\bm{\omega}
\chi_{(\bm{p}_1-\bm{p}_2)\cdot\bm{\omega}>0}
(\bm{p}_1-\bm{p}_2)\cdot\bm{\omega}\times\\
&\left[
\frac{1}{\alpha^2}f^{(2)}(\bm{r}_1,\bm{r}_1-\sigma\bm{\omega},\bm{p}_1,\bm{p}_2,t)
-
f^{(2)}(\bm{r}_1,\bm{r}_1+\sigma\bm{\omega},\bm{p}_1',\bm{p}_2',t)
\right]
\label{eq:CollisionOperator}
\end{align}
where $\bm{p}_i'$ is the pre-collisional velocity associated to $\bm{p}_i$, determined using \cref{eq:CollisionRule}.
Long range interactions are included in
\begin{align}
\mathcal{L}_{\mathrm{part}}(f^{(2)},f^{(3)},...,f^{(N)})=&
\int_{\mathbb{R}^d}\int_{\mathbb{R}^d} \diff\bm{r}_2\diff\bm{p}_2\bm{\nabla_{r}}v_2(\bm{r},\bm{r}_2)\bm{\cdot\nabla_\bm{p}}f^{(2)}\nonumber\\
&-\int_{\mathbb{R}^{2d}}\int_{\mathbb{R}^{2d}} \diff\bm{r}_2\diff\bm{p}_2\diff\bm{r}_3\diff\bm{p}_3\bm{\nabla_{r}}v_3(\bm{r},\bm{r}_2,\bm{r}_3)\bm{\cdot\nabla_\bm{p}}f^{(3)}+...,\label{eq:InterparticleTerm}
\end{align}
where, for example, $v_2(\bm{r},\bm{r}_2)$ relates to the two-body potential $V_2(\bm{r}_i,\bm{r}_j)$, where the prefactor of $\frac{1}{2}$ has been absorbed by a symmetry argument.
\subsection{DDFT Derivations}
As the BBGKY equations are hierarchical, they do not constitute a small enough closed set of equations for efficient simulation. We therefore must truncate the hierarchy, and introduce additional assumptions to close the remaining set of equations.

\Cref{eq:OneBodyKramers} involves integrals with higher order distribution functions which must be approximated. We consider moments of \cref{eq:OneBodyKramers}, and close this system by approximating higher order moments and distributions using lower order counterparts. We first approximate the many-body interactions (\cref{eq:InterparticleTerm}) in the non-equilibrium fluid by those of an equilibrium fluid with the same one body density profile \cite{Archer2009}:
\begin{gather}
\rho(\bm{r})\bm{\nabla_{r}} \frac{\delta\mathcal{F}_{\rm ex}[\rho(\bm{r})]}{\delta\rho(\bm{r})}=-\int_{\mathbb{R}^d}\diff\bm{p}\mathcal{L}_{\text{part}}(f^{(2)},f^{(3)},...,f^{(N)}),\label{eq:ExcessFreeEnergy}
\end{gather}
where $\delta\mathcal{F}_{\rm ex}[\rho(\bm{r})]/\delta\rho(\bm{r})$ is the functional derivative of the excess part of the Helmholtz free energy functional. We also assume that higher order distributions are uncorrelated in velocity:
\begin{align}
f^{(k)}(\bm{r}^k,\bm{p}^k,t)&=g^{(k)}(\bm{r}^k,t)\prod_{i=1}^kf^{(1)}(\bm{r}_i,\bm{p}_i),\label{eq:Nand1Correlations}
\end{align}
for $2\le k\le N$. We note that \cref{eq:CollisionRule} implies there is correlation in velocity, however we expect these correlations to be short range and therefore dominated by correlations in position.

We note that, upon truncation of the hierarchy and approximation of higher order distributions in terms of low order counterparts, information from the Liouville equation (the $N^{\mathrm{th}}$ equation in the BBGKY hierarchy), in particular volume exclusion effects, are lost. It is therefore necessary to include a term which approximates volume exclusion from the Liouville equation. It is popular to include a pairwise interaction `potential' which forbids overlap:
\begin{align}
V_2(\bm{r}_1,\bm{r}_2,t) = \begin{cases}
0, & \text{ if }\|\bm{r}_1-\bm{r}_2\|>\sigma,\\
\infty, & \text{ if }\|\bm{r}_1-\bm{r}_2\|\le\sigma.
\end{cases}
\label{eq:V2Exclusion}
\end{align}

The Helmholtz free energy functional can be generalised to include volume exclusion due to hard particles via a suitable modification of $\mathcal{F}_{\rm ex}$. In one dimension we use the exact functional for volume exclusion derived by Percus \cite{Tarazona2008}, while Fundamental Measure Theory (FMT) is used to accurately approximate volume exclusion for spheres \cite{Rosenfeld1989} for $d>1$. However, neither the `potential' in \eqref{eq:V2Exclusion} nor FMT directly include the effect of collisions in the system, which must be included using a collision operator.

The collision operator is also an alternative way of including volume exclusion effects. In \cite{Marini2007}, in one dimension, volume exclusion effects are incorporated by approximating the correlation function $g^{(2)}$ in the RET collision operator using an analytic form:
\begin{align}
g^{(2)}(r,r\pm\sigma) = \frac{1}{1-\eta(r\pm\sigma/2)},\label{eq:gRET}
\end{align}	
where $\eta(x)$ is the local packing fraction. As the packing fraction approaches 1, the value of $g$ given by \cref{eq:gRET} blows up and, in an analogous way to the Percus free energy in \cref{eq:Percus}, causes volume exclusion in the model. However, the numerics in the present work show that this approximation is not accurate for dynamics with inelastic collisions, where, in particular, the value of the correlation function at contact increases for low local densities, rather than decreasing as in the elastic case. In \Cref{Section:Numerics} we provide an example which shows that when the correlation function is approximated by experimental data and a volume exclusion free energy term is absent, the local density can exceed physical limits (see \cref{Figure:rhoNoPercus}).

Returning to the derivation of a continuum model, we define $\rho,\bm{v}$ and $\bm{E}$ as the number density, the local average velocity and the granular temperature of the system respectively:
\begin{gather}
\rho(\bm{r},t) =
\int_{\mathbb{R}^{d}}\diff\bm{p}f^{(1)}(\bm{r},\bm{p},t), \quad \bm{v}(\bm{r},t) = \frac{1}{\rho(\bm{r},t)}
\int_{\mathbb{R}^{d}}\diff\bm{p}\frac{\bm{p}}{m}f^{(1)}(\bm{r},\bm{p},t)
,\nonumber\\
\bm{E}(\bm{r},t) =
\int_{\mathbb{R}^{d}}\diff\bm{p}\frac{\bm{p}\otimes\bm{p}}{m^2}f^{(1)}(\bm{r},\bm{p},t)
.
\end{gather}
We finally assume that the one particle distribution function can be approximated by a local equilibrium Maxwell-Boltzmann distribution \cite{Hansen1990}:
\begin{align}
f^{(1)}_{le}(\bm{r},\bm{p},t)&=\frac{\rho(\bm{r},t)}{|2\pi mk_BT\bm{E}(\bm{r},t)|^{1/2}}\exp\left(-\frac{(\bm{p}-m{\bm{v}}(\bm{r},t))^T\bm{E}(\bm{r},t)^{-1}(\bm{p}-m\bm{v}(\bm{r},t))}{2mk_BT}\right).\label{eq:MaxBoltzLocalEq}
\end{align}
It is well known that the local equilibrium of a granular fluid is in fact not Maxwellian \cite{Benedetto1998}, and other approximations are a topic of current research \cite{Garzo2007}. However, the assumption \cref{eq:MaxBoltzLocalEq} allows us to write the second moment of $f^{(1)}(\bm{r},\bm{p},t)$ as a product of the density $\rho(\bm{r},t)$ and local average velocity $\bm{v}(\bm{r},t)$. We expect other functional forms of the local equilibrium can be implemented in the same manner, but to introduce the model we use \cref{eq:MaxBoltzLocalEq}.

The continuity equation is then derived by integrating \cref{eq:OneBodyKramers} with respect to $\bm{p}$, under the assumptions stated:
\begin{align}
\frac{\partial\rho}{\partial t}= -\bm{\nabla_{r}\cdot}(\rho\bm{v})\label{eq:ContinuityDDFT}.
\end{align}
By multiplying \cref{eq:OneBodyKramers} by $\bm{p}$, then integrating with respect to $\bm{p}$, standard calculus results lead to the momentum equation \cite{Archer2009}, which now includes the granular temperature, and the first moment of the collision operator:
\begin{align}
\frac{\partial \bm{v}}{\partial t}
+\bm{v\cdot\nabla_{r}v}
+\gamma\bm{v}
+ \frac{k_BT}{m\rho}\bm{\nabla_{r}\cdot}\left(\rho\left(\bm{E}-\bm{I}\right)\right)
+\frac{1}{m}\bm{\nabla_{r}}\frac{\delta \mathcal{F}[\rho]}{\delta\rho}
-\frac{1}{m\rho}\bm{\mathcal{M}}_1(\mathcal{L}_{\mathrm{coll}})
=0,\label{eq:MomentumDDFT}
\end{align}
and also includes the Helmholtz free energy functional:
\begin{align}
&\mathcal{F}[\rho]\coloneqq k_BT\int \diff\bm{r}\rho(\bm{r})[\ln\Lambda^3\rho(\bm{r},t)-1]+\mathcal{F}_{\rm ex}[\rho(\bm{r},t)]+\int \diff\bm{r}V_{ext}(\bm{r})\rho(\bm{r},t),\label{eq:HelmholtzFE}
\end{align}
where $\Lambda$ is the (irrelevant) {\em thermal de Broglie wavelength}. Finally, when considering the third moment by multiplying by $\bm{p}\otimes\bm{p}$ then integrating with respect to momentum, by using \cref{eq:MomentumDDFT} and \cref{eq:ContinuityDDFT}, we arrive at an equation describing the evolution of the granular temperature:
\begin{align}
\partial_t\bm{E} 
+ \bm{v\cdot\nabla_{r} E} + (\bm{E \nabla_{r} v}) + (\bm{E \nabla v})^T
+2\gamma(\bm{E} - \bm{I})
-\frac{1}{k_BT\rho}\bm{\mathcal{M}}_2(\mathcal{L}_{\mathrm{coll}})
=0.\label{eq:GranularTempDDFT}
\end{align}
\Cref{eq:MomentumDDFT,eq:GranularTempDDFT} include centred moments of the collision operator:
\begin{align}
\bm{\mathcal{M}}_1(\mathcal{L}_{\mathrm{coll}}) =& \int_{\mathbb{R}^d}\diff\bm{p}\frac{(\bm{p}-\bar{\bm{p}})}{m}\mathcal{L}_{\text{coll}}(f^{(1)},g^{(2)}),\label{eq:LcollMoment1}\\
\bm{\mathcal{M}}_2(\mathcal{L}_{\mathrm{coll}}) =&
\int_{\mathbb{R}^d}\diff \bm{p}\frac{(\bm{p}-\bm{\bar{p}})\otimes(\bm{p}-\bm{\bar{p}})}{m^2}\mathcal{L}_{\text{coll}}(f^{(1)},g^{(2)}),\label{eq:LcollMoment2}
\end{align}
where the argument of $\mathcal{L}_{\mathrm{coll}}$ has changed to account for the assumption \cref{eq:Nand1Correlations}. For \cref{eq:CollisionOperator}, by applying \cref{eq:MaxBoltzLocalEq}, we can write \cref{eq:LcollMoment1,eq:LcollMoment2} in terms of Gaussians and error functions. The exact forms used in simulations are given in \cref{Appendix:CollisionOperator}.

Given the centred moments of $\mathcal{L}_{\text{coll}}(f^{(1)},g^{(2)})$, and the correlations $g^{(k)}$, the set of equations \cref{eq:ContinuityDDFT,eq:MomentumDDFT,eq:GranularTempDDFT} then constitute a closed model for granular media, incorporating volume exclusion due to hard particles, external and inter-particle potentials, and (in)elastic collisions. 

There are some important differences between the DDFT model here and existing DDFTs in the literature. Firstly, we include moments of the collision operator \cref{eq:LcollMoment1,eq:LcollMoment2}, which must be included to incorporate dissipative effects due to inelastic collisions. Our numerical experiments (see \cref{Figure:RadialCorrelations}) show that the collision terms do affect the dynamics, and that the effect is pivotal when $\alpha<1$. We also include an additional moment \cref{eq:GranularTempDDFT} of \cref{eq:OneBodyKramers}, as the effects of the collision term are evident in the granular temperature of the system; \cref{eq:CollisionRule} reduces the variance of particle velocities when $\alpha<1$, so we expect it to have a dissipative effect on $\bm{E}(\bm{r},t)$. In particular under our assumptions, in one dimension,
\begin{align}
\lim_{\sigma\rightarrow0}\bm{\mathcal{M}}_1(\mathcal{L}_{\mathrm{coll}}) =0,\quad
\lim_{\sigma\rightarrow0}\bm{\mathcal{M}}_2(\mathcal{L}_{\mathrm{coll}}) =- \frac{2g^{(2)}(r)\rho^2(mk_BTE)^{3/2}(1-\alpha^2)}{\sqrt{\pi}}.
\end{align}
Thus inelasticity has a small effect on \cref{eq:MomentumDDFT}, but can be incorporated by including an additional moment.

Although DDFT derivations involving collision terms \cite{Marini2007} and temperature gradients \cite{Wittkowski2012} have been studied, it is clear that for granular media both terms play important roles. When comparing to results in kinetic theory, the addition of the free energy term allows us to include effects both from interparticle interactions and volume exclusion, by considering the interactions at the particle level.

We also note the importance in the choice of initial condition; the initial density, velocity and granular temperature must satisfy the physical restrictions of the PDE; in one dimension this corresponds to not exceeding the packing fraction limit $\rho_v=1$, and that the density must be non-negative $\rho(r)\geq 0$ for all $r$.

In one dimension, we can incorporate volume exclusion exactly using the Percus free energy \cite{Percus1976}:
\begin{align}
\frac{\delta \mathcal{F}_{\mathrm{ex}}[\rho]}{\delta\rho}(x) = 
\log\left(
1 - \int_{x}^{x+\sigma}\rho(x')\diff x'
\right)
+ \int_{x-\sigma}^{x}
\frac{\rho(x'')}{1-\int_{x''}^{x''+\sigma}\rho(x')\diff x'}\diff x''\label{eq:Percus}
\end{align}
We note that in one dimension for constant $\rho$, as $\rho\rightarrow\frac{1}{\sigma}$, we approach the maximum density and so $\frac{\delta \mathcal{F}_{\mathrm{ex}}[\rho]}{\delta\rho}\rightarrow\infty$, {\em i.e.} the chemical potential of the system blows up. This can be seen as a constraint on adding particles to the system, and implicitly stops the volume of the system increasing beyond the physical limit; \cref{eq:Percus} is defined so that it is limited by the close packing value. Thes effects are not present when solely including a collision operator with arbitrary $g$, such as those obtained from microscopic simulations.

\section{Numerical results}\label{Section:Numerics}
\subsection{Parametrisation of $g^{(2)}(\sigma)$}
To use \cref{eq:ContinuityDDFT,eq:MomentumDDFT,eq:GranularTempDDFT} derived in \Cref{Section:Derivation} we need accurate approximations for the functions $g^{(k)}$, in particular the pair correlation function $g^{(2)}$, if we assume that particle interactions are pairwise. Analytical approaches to finding $g^{(2)}$ are generally restricted to simple systems \cite{Ibsen1997}. One can construct an additional coupled $2d$-dimensional PDE for $g^{(2)}$ via moment closure schemes \cite{Hughes2012}, but this increases the dimensionality of the problem and so introduces a prohibitively larger computational cost when $d>1$.

\begin{figure}
	\centering
	\subfloat[][]{\includegraphics[width=0.45\textwidth]{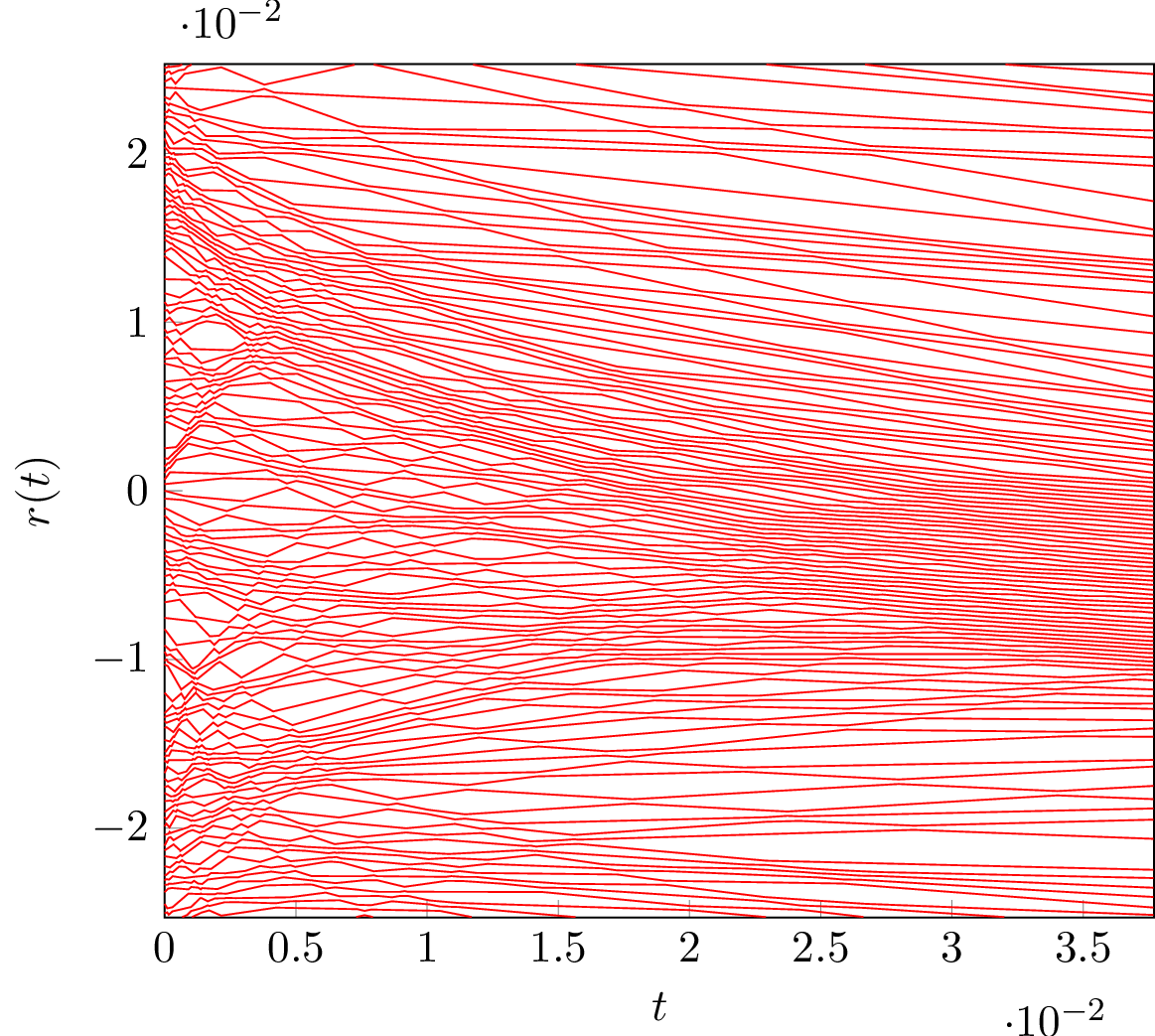}\label{Figure:SampleAlpha0.5}}
	\subfloat[][]{\includegraphics[width=0.45\textwidth]{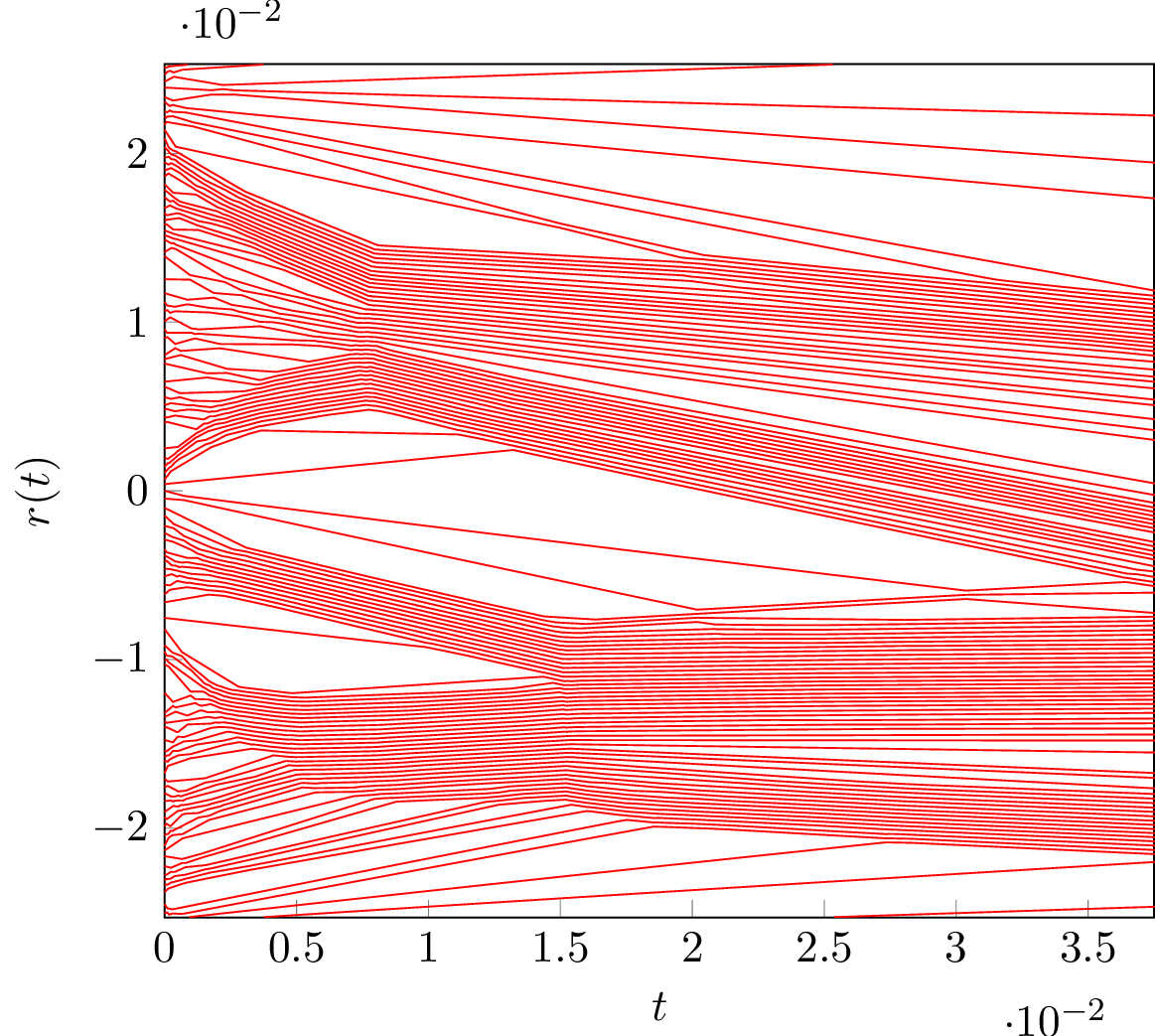}\label{Figure:SampleAlpha0.9}}\\
	\caption{Samples that can be used to construct $g^{(2)}(\sigma)$ at different points in time. The red lines are trajectories of the centres of mass of individual particles. We note that the domain is periodic, so trajectories can disappear and reappear at the top and bottom of the $y$ axis. Left: $\alpha=0.9$, right: $\alpha=0.5$. In these examples the initial conditions are the same, but display characteristic differences in their dynamics, in particular when $\alpha=0.5$ particle streaming is more evident.}\label{Figure:SampleExamples}
\end{figure}

It is known \cite{Rex2009} that higher order correlations equilibriate much faster than the density. This result validates the {\em adiabatic approximation}; correlations can be approximated by their local equilibrium values. Many empirical forms for $g^{(2)}$ at equilibrium (or quasi-equilibrium) are constructed and used in the literature, although it is reasonable to expect that $g^{(2)}$ depends on properties of the particles and dynamics. In particular, inelasticity leads to the effect of particle streaming, which should be visible in the correlation function (see \cref{Figure:SampleExamples}).

We therefore empirically construct $g^{(2)}$ for the system of interest. By applying statistical methods on synthetic data generated by extensive particle simulations with small $N$, we parametrise $g^{(2)}$ without simulating the entire system, avoiding excessive computational cost. We can then incorporate it in the model \cref{eq:ContinuityDDFT,eq:MomentumDDFT,eq:GranularTempDDFT}. Analogously, parametrisations could be performed with experimental data.

We present an example of this methodology using a system of 100 deterministic hard rods ($d=1$) on a $2\pi$-periodic domain in the absence of external or interparticle potentials, with $\gamma=2$ and $\sigma=2\pi/25,000$. In the absence of a thermostat, the trajectories of individual particles can then be solved analytically. Therefore, instead of a numerical method involving a timestep for microscopic simulation, we predict collision times of particles, then sort and schedule and process these events before advancing simulation. This methodology is at the centre of Event Driven Particle Dynamics (EDPD), which was first developed as early as the 1950s \cite{Alder1959}, but is still a modern topic of research.

A naive EDPD algorithm will predict future collisions between all particles after each collision has been processed, producing an algorithm with computational complexity $O(N^2)$ per collision. Our simulations use the cell method \cite{Alder1957} to reduce the computational cost of event prediction to $O(N)$. We also update the position of each particle asynchronously \cite{Lubachevsky1991}, which reduces the cost of advancing simulation to $O(1)$. In addition, data structures such as binary search trees \cite{Rapaport1980} and bounded increasing priority queues \cite{Paul2007} can be implemented to decrease the computational cost of event sorting and scheduling to $O(1)$ per collision. Combining methods for prediction and scheduling gives dynamics with a cost of $O(1)$ per collision. Software packages such as \texttt{DynamO} \cite{Bannerman2011} implement all of these methods, but are currently limited to systems with friction coefficient $\gamma=0$. Finally, methods to efficiently parallelise EDPD algorithms have also been constructed \cite{Herbordt2009}. 

\begin{figure}
	\centering
	\subfloat[][]{\includegraphics[width=0.45\textwidth]{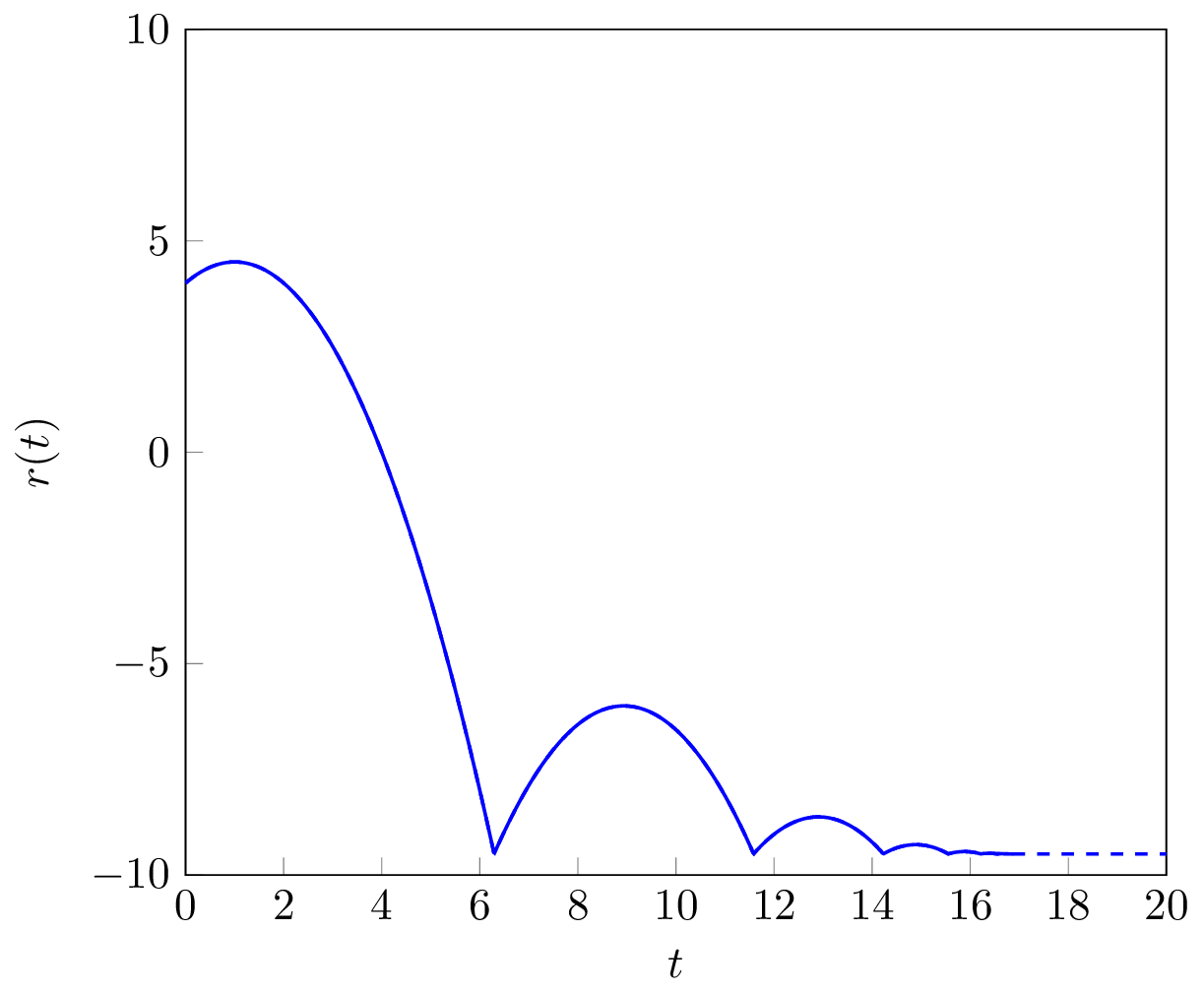}\label{Figure:GravityInelasticCollapse}}
	\subfloat[][]{\includegraphics[width=0.45\textwidth]{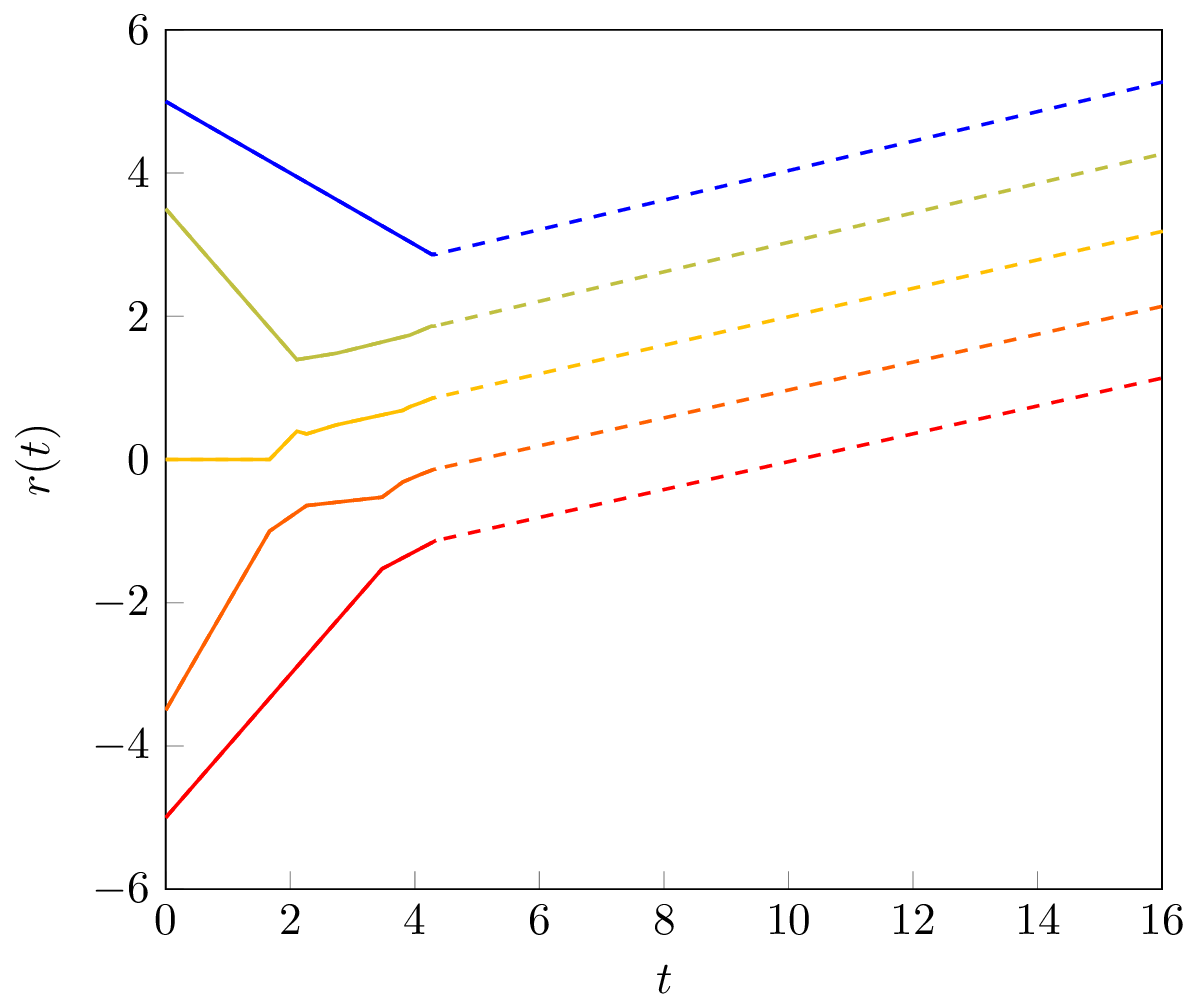}\label{Figure:InelasticCollapse}}\\
	\caption{EDPD simulations displaying inelastic collapse in one dimension. Left: A single inelastic particle with $\alpha=0.8$ under the effect of gravity in a domain with hard walls. Right: A collection of inelastic particles with $\alpha=0.5$ in a periodic domain. The solid lines and dashed lines show the same number of collisions, where the TC method has been implemented for the dashed lines with $t_c=10^{-5}$, allowing dynamics to advance.}\label{Figure:InelasticCollapseExample}
\end{figure}

In addition, several numerical methods are available to avoid inelastic collapse in EDPD. A review of these methods is available in \cite{Luding1998}, and in our simulations, we implement the TC model, which renders collisions elastic when a particle undergoes a collision a small time $t_c$ after a previous collision. Together with polydispersity, the TC method stops sharp peaks from forming in the radial correlation function, but also accurately approximates the dynamics of the system. We give two examples of situations that undergo inelastic collapse in one dimension in \cref{Figure:InelasticCollapseExample}. When $\alpha<1$, an infinite number of collisions occurs in finite time, so the dynamics are `jammed'. The TC model allows the particle to `vibrate' when the collisions become very frequent.
We note that, by design, volume exclusion is incorporated in the algorithm; collisions are predicted and processed so that particles do not overlap. Therefore, unlike in continuum modelling, we do not require any additional potential in these dynamics to include volume exclusion effects.

Using EDPD to construct $5000$ samples of each system with a range of values of $\alpha$ and solid volume fractions $\rho_v$, we construct a parametrisation of $g^{(2)}(\sigma)$. Examples of systems with $\alpha=0.9$ and $\alpha = 0.5$ are given in 
\cref{Figure:SampleExamples}.
For $d=1$ in the absence of friction, external and interparticle potentials, the radial correlation function blows up when particles are monodisperse and inelastic; at equilibrium all particles will be moving in contact. We therefore include a variance in the diameter of particles $\sigma_v = 0.1\sigma$. \Cref{eq:ContinuityDDFT,eq:MomentumDDFT,eq:GranularTempDDFT} can be adapted to take into account poly-dispersity, but we expect the effect to be negligible in this case and so we ignore it in the DDFT. In each sample, the initial velocities of the particles are normally distributed with mean $0$ and variance $1$, and positions are drawn from a uniform distribution in the domain. We evolve the  system until $99.9\%$ of the energy of the particles has dissipated due to inelasticity and friction, then construct a near equilibrium parametrisation of $g^{(2)}(\|r_1-r_2\|)$. We note that when $\rho_v$ is small and $\alpha$ is close to 1, few collisions happen before the effect of friction causes particles to lose all their energy, and when $\rho_v=1$, the system is fully dense, so the value of the correlation function is independent of $\alpha$.

\begin{figure}
	\centering
	\subfloat[][]{\includegraphics[width=0.45\textwidth]{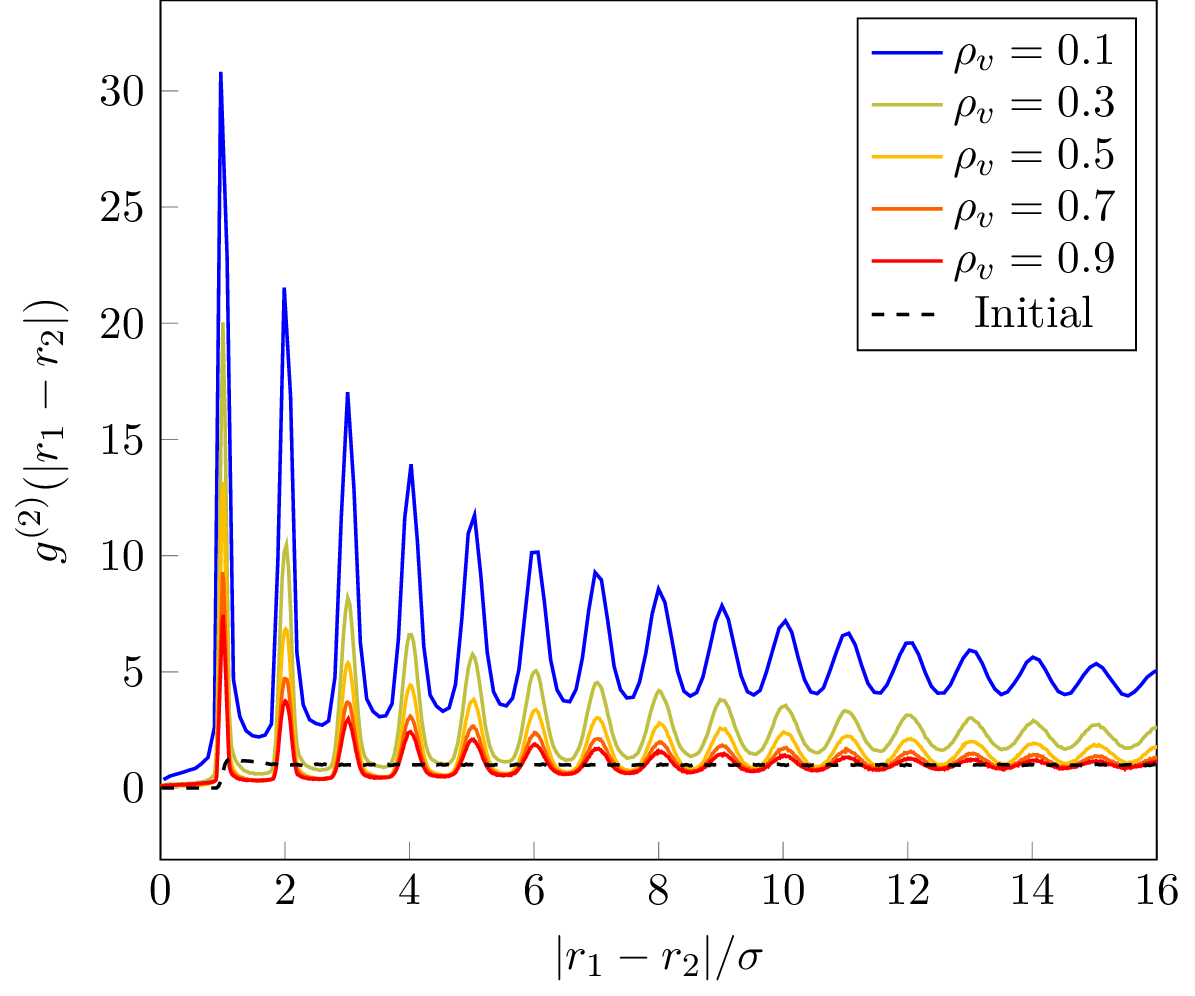}\label{Figure:RadialCorrelations}}
	\subfloat[][]{\includegraphics[width=0.45\textwidth]{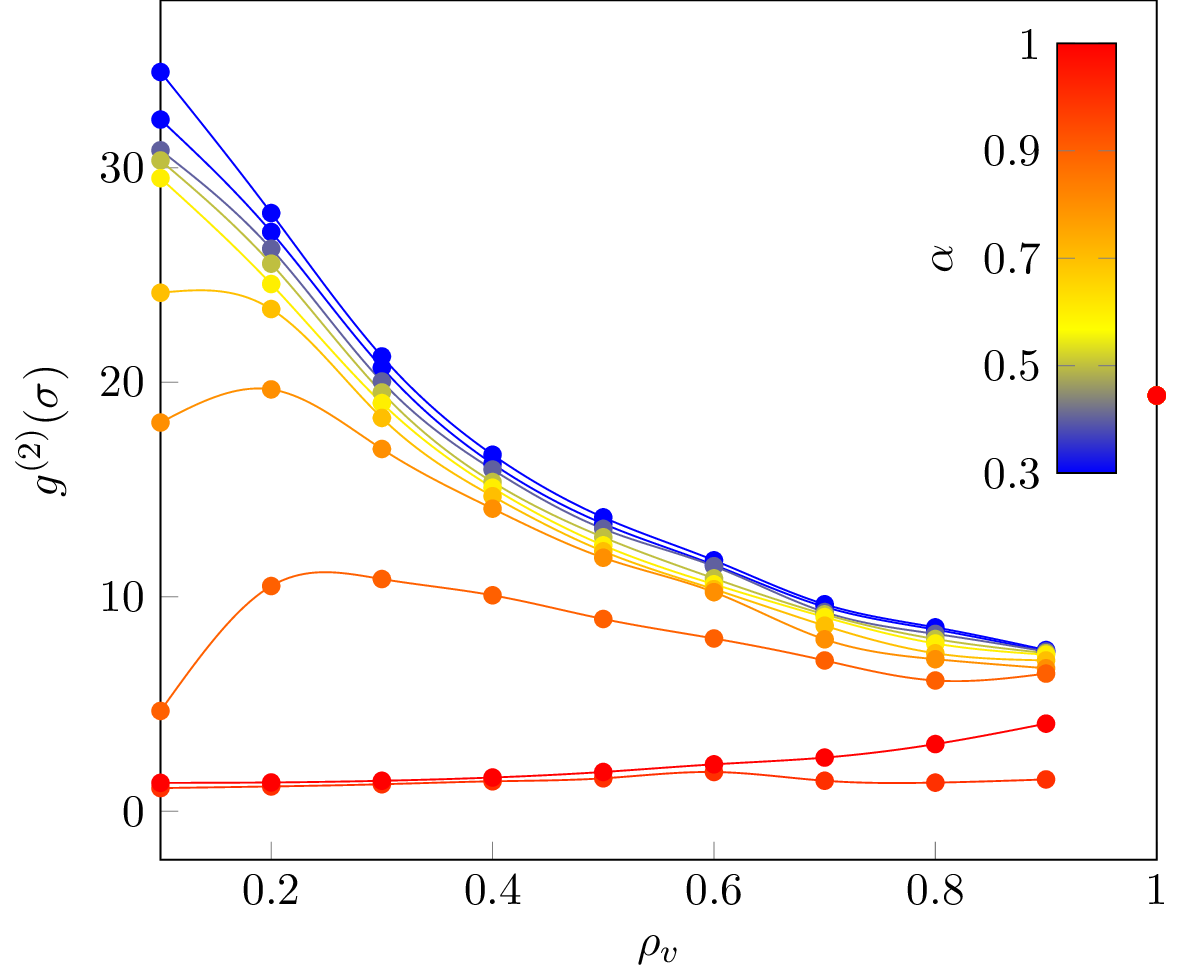}\label{Figure:RadialFunction}}\\
	\caption{Left: the radial correlation function $g^{(2)}(\|r_1-r_2\|)$ for $\alpha=0.5$, with different solid fractions $\rho_v$. Right: cubic spline interpolation of $g^{(2)}(\sigma)$ varying over $\rho_v$, for different values of $\alpha$.}		
\end{figure}

In \cref{Figure:RadialCorrelations}, we display $g^{(2)}(\|r_1-r_2\|)$ for different densities, as well as $g$ at time $t=0$. We note that for low densities and $\alpha=0.5$ the radial correlation function has several peaks. This is evidence of particle streaming, where inelastic collisions cause particles to move at the same velocity, near to one another.

Under the assumption that $g^{(2)} = g^{(2)}(\|r_1-r_2\|)$, only the value of $g^{(2)}$ at $\|r_1-r_2\|=\sigma$ is included in the collision terms. In \cref{Figure:RadialFunction} we use the results of particle simulations to parametrise $g^{(2)}(\sigma)$ for different $\alpha$ and $\rho_v$, using cubic smoothing spline interpolation \cite{Craven1978}. We omit the point when $\sigma_v=1$ from the interpolation to improve the curve fit for values used in the DDFT simulation.
\subsection{DDFT simulation}
For a continuum approach we simulate \cref{eq:ContinuityDDFT,eq:MomentumDDFT,eq:GranularTempDDFT}. We use pseudospectral code provided in \cite{Nold2017}, which is available at \cite{DDFTCode}. We consider a periodic domain $[0,100]$ with $100$ computational points. Using more than 100 points has little effect on the result of the simulation. To match microscopic simulations, we set $\gamma=2$ and $\sigma=2\pi/25000$, so that there are $120,000$ particles in the system (a fully packed domain would hold $400,000$ particles in this case). We include the Percus free energy functional \cite{Percus1976} to incorporate volume exclusion in the dynamics. 

Before continuing, we note that the system can be made dimensionless by considering the following scaling:
\begin{align}
\gamma\sim\frac{1}{T}, \quad 
k_BT\sim \frac{L^2M}{T^2}, \quad
\rho\sim\frac{1}{L},\quad
v\sim\frac{L}{T}, \quad
m\sim M,
\end{align}
where $L$ is a length scale (in our simulations we use the domain length, but the particle diameter could also be considered), $T$ is a time scale and $M$ is a mass scale (the mass of a particle). Furthermore, the granular temperature is dimensionless. The scaling of variables and parameters are constructed by considering the the microscopic dynamics and the distributions $f^{(N)}$ as in \cite{Marini2007}.

The initial conditions are given by
\begin{gather}
\rho_0(r) = \frac{\rho_v}{N_c}\left(e^{\frac{(r-25)^2}{25}} + e^{\frac{(r-75)^2}{25}}+0.5\right),\quad
v_0(r)    = 20\sin\left(\frac{2\pi r}{100}\right),\quad
E_0(r)    = 250,\label{eq:InitialConditions}
\end{gather}
where $N_c$ is a normalisation constant, and $\rho_v=0.3$ is the total solid volume fraction. The initial conditions are chosen such that areas of higher density will move toward one another and `collide'. 

\begin{figure}
	\centering
	\includegraphics[width=\textwidth]{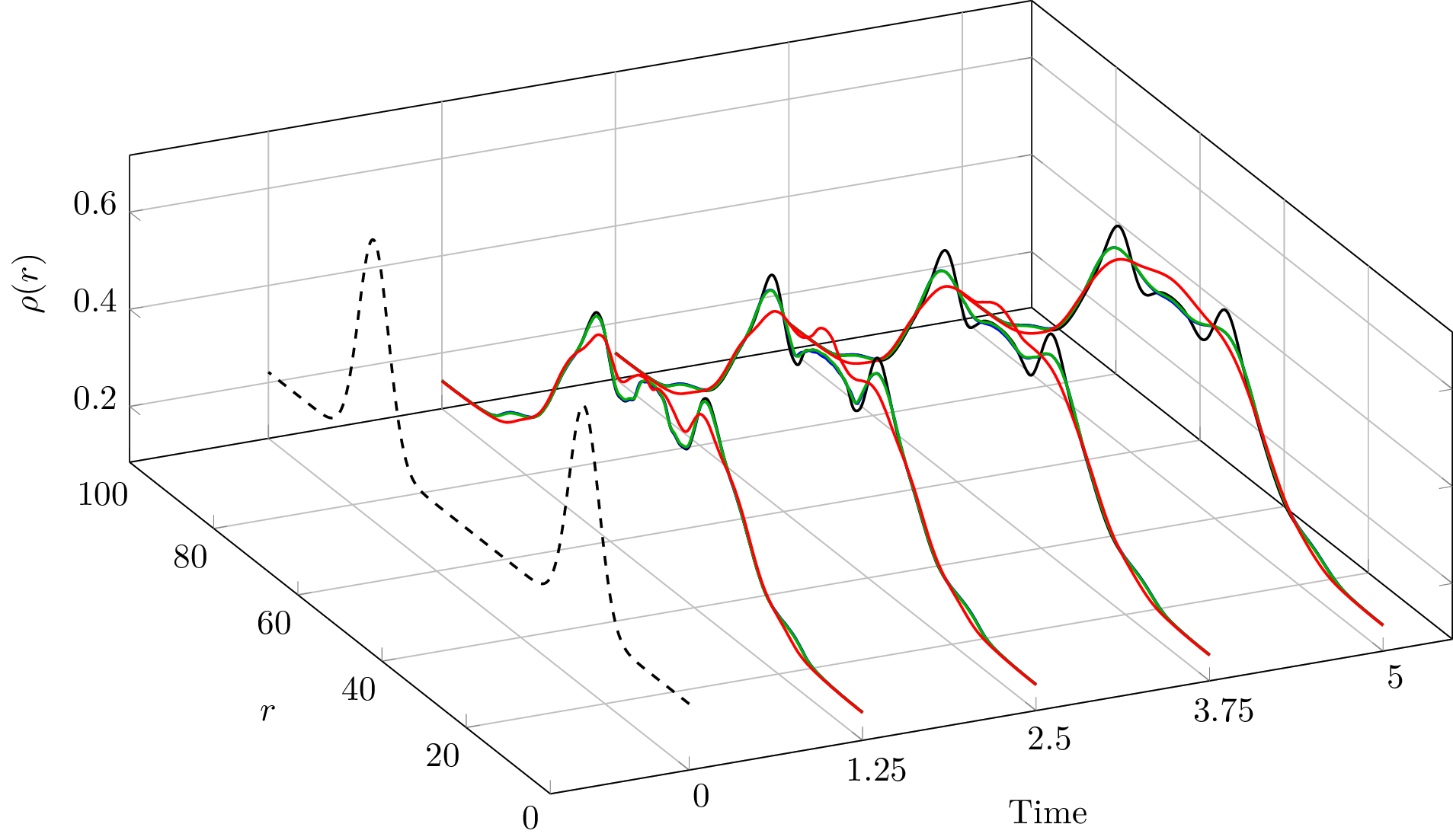}
	\caption{Results from the DDFT simulation for $\alpha=0.5$. Each simulation has the same initial condition (black, dashed) at time $0$.
		The black line neglects the collision operator and the free energy term. The blue line includes the free energy term but not the collision operator. The green line includes both terms, with $g^{(2)}(\sigma)=1$, and the red line includes both terms and uses $g^{(2)}(\sigma)$ determined by particle simulations, shown in \cref{Figure:RadialFunction}.}\label{Figure:rho}
\end{figure}

\begin{figure}
	\centering
	\includegraphics[width=\textwidth]{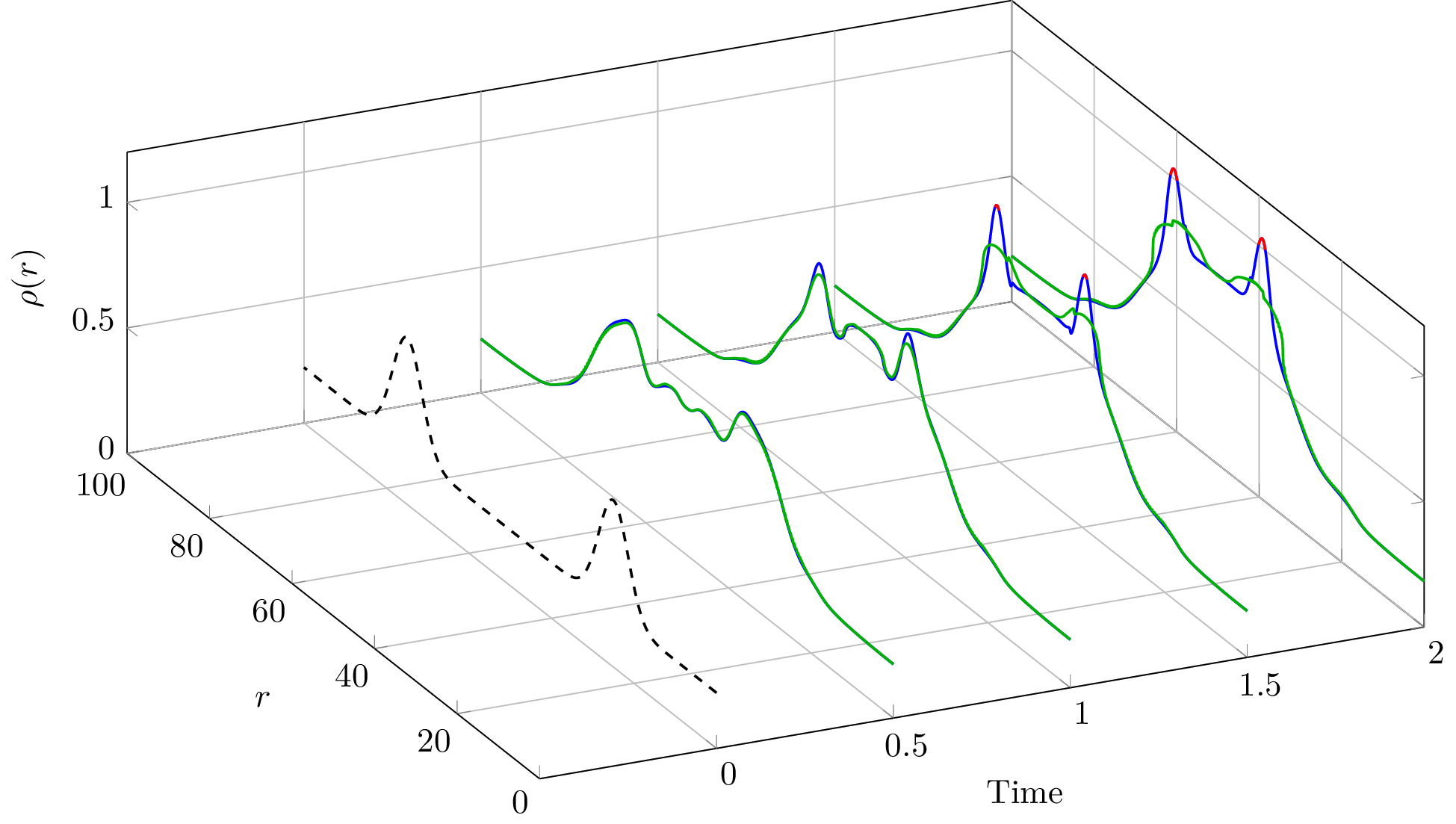}
	\caption{Results from the DDFT simulation, using $g^{(2)}(\sigma)$ determined by particle simulations for $\alpha=1$, and the same initial condition (black, dashed). The blue line gives the result when the free energy term is neglected, and the green is the same simulation with the Percus free energy term included. Any values of $\rho$ which are above the physical limit of $\rho=1$ are coloured red. In this example we used 600 computational points, to ensure that the volume exclusion effects are numerically stable.}\label{Figure:rhoNoPercus}
\end{figure}

\begin{figure}
	\centering
	\includegraphics[width=\textwidth]{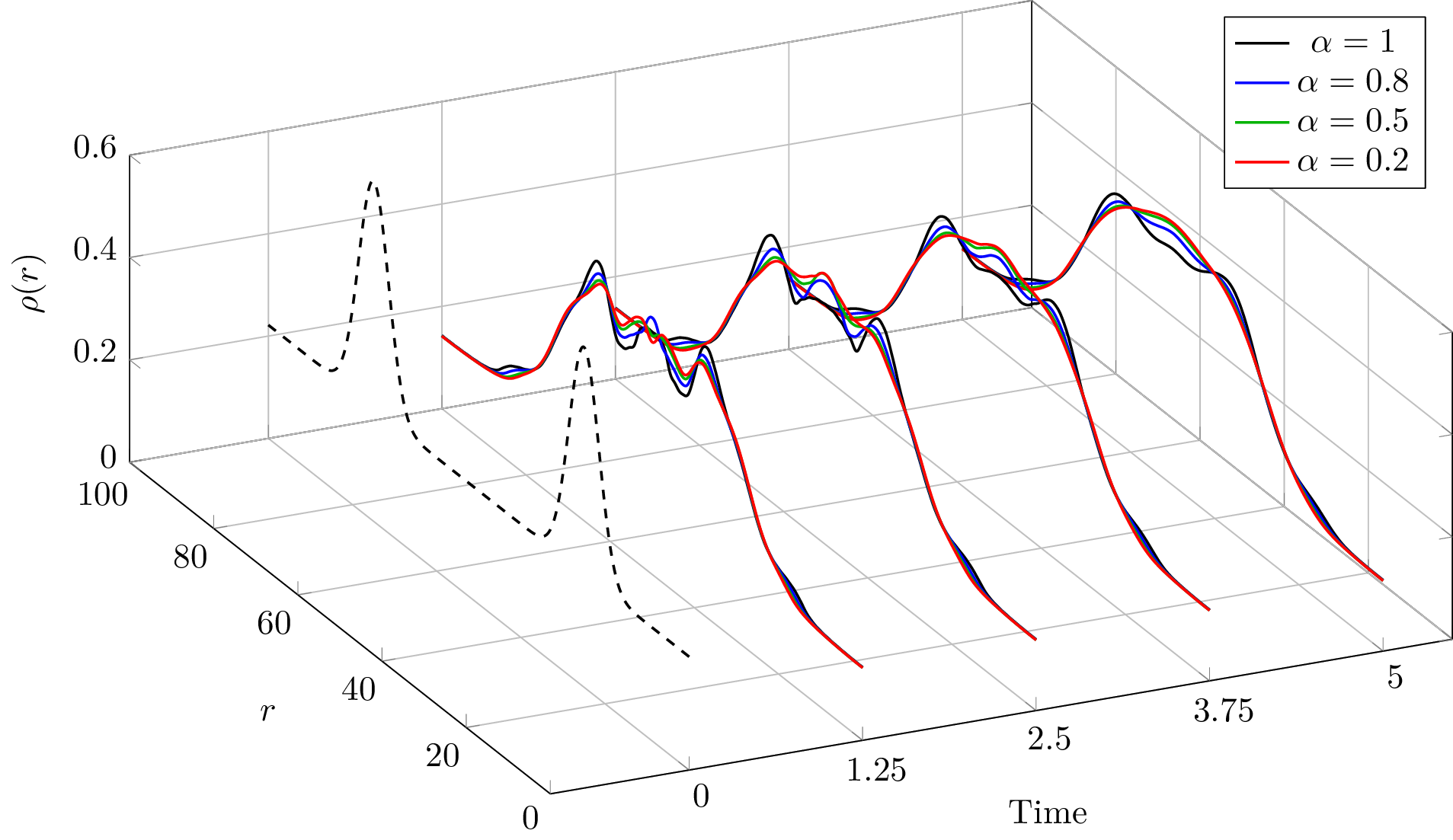}
	\caption{Results from the DDFT simulation, using $g^{(2)}(\sigma)$ determined by particle simulations for different $\alpha$, using the same initial condition (black, dashed).}\label{Figure:rhoDifferentAlpha}
\end{figure}

\begin{figure}
	\centering
	\includegraphics[width=0.45\textwidth]{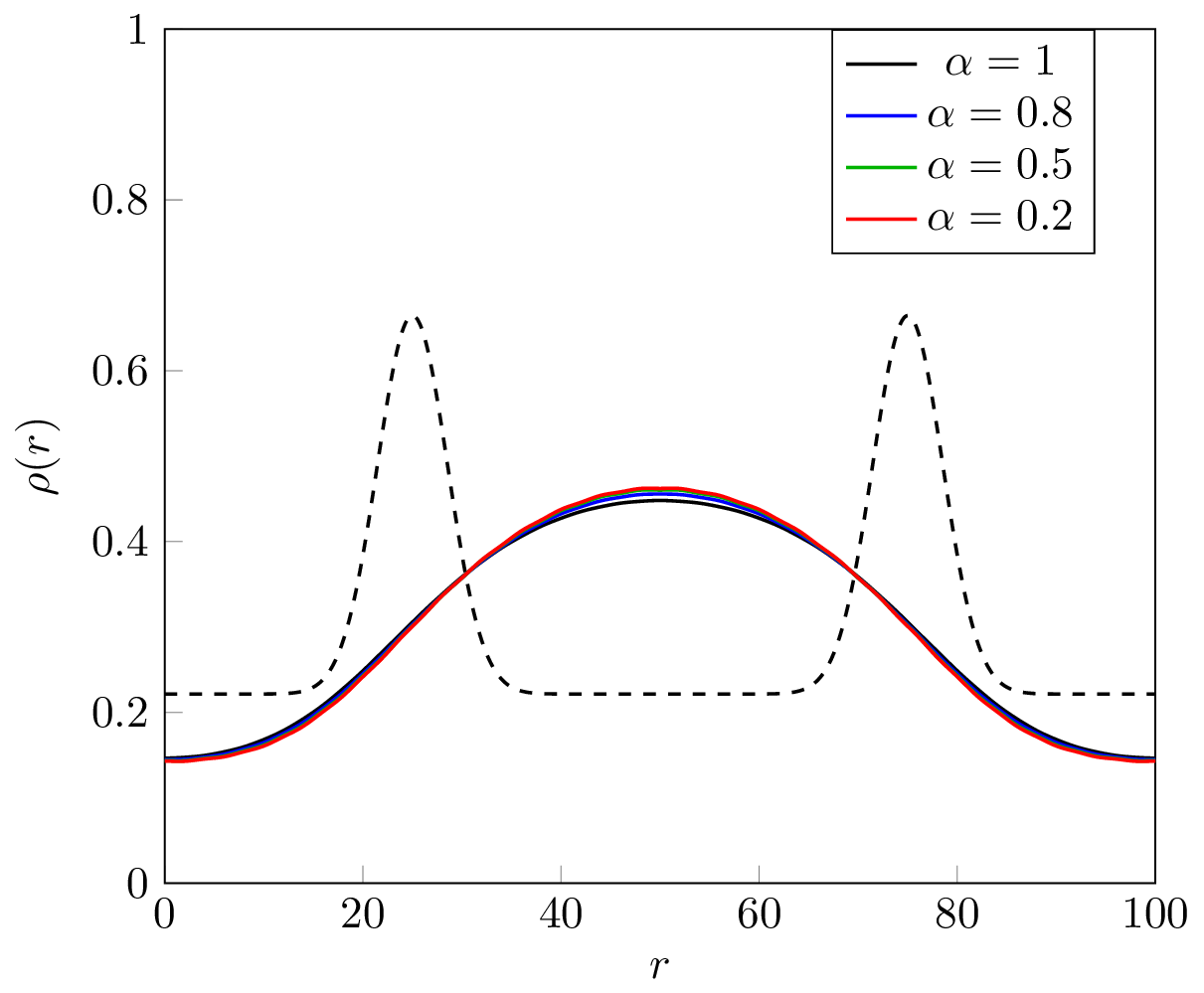}\\
	\caption{The density at time $T=50$ for different coefficients of restitution, compared to the initial density (black, dashed).}\label{Figure:EquilibriumDetails}		
\end{figure}

\Cref{Figure:rho} displays results of model \cref{eq:ContinuityDDFT,eq:MomentumDDFT,eq:GranularTempDDFT} at different times $t$ when $\alpha=0.5$. The results show that every term is necessary for accurate dynamics: If the Percus term is neglected the density is sometimes overestimated as particle volume exclusion of hard particles is not incorporated. If the collision term is included but $g^{(2)}(\sigma)=1$ (i.e. the uncorrelated case) the  inelastic effects are not noticeable, and the high density areas are reflected upon `collision'. When all terms are present with $g^{(2)}$ constructed from particle simulations we see that the two higher density areas coalesce, an intuitive result for inelastic dynamics.
To show the importance of including a volume exclusion free energy term in the DDFT, we consider an example with modified initial conditions: we set $N=175,000$, and
\begin{align}
v_0(r) = 26 \sin\left(\frac{2\pi r}{100}\right).
\end{align}
The results in \cref{Figure:rhoNoPercus} show that the system reached an unphysical density in finite time if the Percus free energy is not included.
Finally, we perform the same dynamics for different $\alpha$ with initial conditions \cref{eq:InitialConditions}. 
The results in \cref{Figure:rhoDifferentAlpha} show that particles coalesce more for smaller $\alpha$.
In \cref{Figure:EquilibriumDetails} we provide the density near equilibrium for different coefficients of restitution. We note that in this example, the long time behaviour of the density is similar for all coefficients of restitution. This is because the effect of the collision operator is small when the local average velocity is small, so in this example where energy is not added into the system using any other external potentials, the effect of friction determines the dynamics for long times. 

\section{Conclusions and future work}\label{Section:Conclusions}
We have constructed a new model for granular media, which can incorporate inelastic collisions using classical collision operators, and interparticle interactions using DDFT methods. We have presented a simple example which displays the importance of each term in the model, but the model can also be used for systems in 2D or 3D with more complicated dynamics such as adhesion between particles. Our results show that our methodology is successful; small scale, inexpensive particle dynamics can be used to fine-tune parameters in the mesoscopic model, such as the radial correlation function.

Much of the current research on DDFT for complex fluids can be adapted to the system of equations in this work, including extension to poly-disperse or multi-species systems \cite{Goddard2013}, and inclusion of more complicated drag forces due to interactions between the particles and fluid in the system using a hydrodynamic interaction tensor \cite{Goddard2012}. Further work on fundamental derivations in the style of \cite{Wilkinson2018} will be beneficial to construct collision operators from more complicated dynamics or particles. Furthermore, inclusion of a more accurate local (non-Maxwellian) equilibrium approximation for granular media could improve the model.

The synthetic data presented is an example of how modern computational and data-scientific methods can be applied to fine-tune parameters in continuum models; we parametrise using statistics from  particle simulations. For systems with more complicated interactions between particles we will need to use state of the art particle simulation methods, but our modelling approach avoids the computational bottleneck caused by simulating large numbers of particles.

\subsection{Acknowledgements}
TDH was supported by The Maxwell Institute Graduate School in Analysis and its Applications, a Centre for Doctoral Training funded by the UK EPSRC (EP/L016508/01), the Scottish Funding Council, Heriot-Watt University and the University of Edinburgh. RO and BDG acknowledge the support of EPSRC (EP/N034066/1 and EP/L025159/1, respectively). The authors would like to acknowledge helpful discussions with Dr. M. Wilkinson (Heriot-Watt University).

\bibliographystyle{ieeetr}      
\bibliography{DDFTforGranularMedia}

\appendix
\section{Moments of the collision operator}\label{Appendix:CollisionOperator}
The moments \cref{eq:LcollMoment1,eq:LcollMoment2} can be constructed analytically by using standard results for moments of Gaussians with mean $\mu$ and variance $\varsigma$:
\begin{gather*}
\int_{-\infty}^{\infty} \exp\left(-\frac{(x-\mu)^2}{\varsigma}\right)\diff x 
=
\sqrt{\pi}\sqrt{\varsigma},\quad
\int_{-\infty}^{\infty}x\exp\left(-\frac{(x-\mu)^2}{\varsigma}\right)\diff x  
=
\sqrt{\pi\varsigma}\mu,\\
\int_{-\infty}^{\infty}x^2\exp\left(-\frac{(x-\mu)^2}{\varsigma}\right)\diff x  
=\frac{\sqrt{\pi\varsigma}}{2}(\varsigma+2\mu^2),
\end{gather*}
as well as the following identities for integrals of Gaussians over half-infinite domains:
\begin{gather*}
\int_{-\infty}^\infty \chi_{\pm x>0}\exp\left(-\frac{(x-\mu)^2}{\varsigma}\right)\diff x = \frac{\sqrt{\varsigma\pi}}{2}\left(1\pm\mathrm{erf}\left(\frac{\mu}{\sqrt{\varsigma}}\right)\right),
\\ 
\int_{-\infty}^\infty \chi_{\pm x>0}x\exp\left(-\frac{(x-\mu)^2}{\varsigma}\right)\diff x =
\frac{\sqrt{\varsigma\pi}m}{2}\left(1\pm\mathrm{erf}\left(\frac{\mu}{\sqrt{\varsigma}}\right)\right)
\pm
\frac{\varsigma}{2}\exp\left(-\frac{\mu^2}{\varsigma}\right),
\\
\int_{-\infty}^\infty \chi_{\pm x>0}x^2\exp\left(-\frac{(x-\mu)^2}{\varsigma}\right)\diff x =
\frac{\sqrt{\pi\varsigma}}{2}\left(1\pm\mathrm{erf}\left(\frac{\mu}{\sqrt{\varsigma}}\right)\right)\left(\frac{\varsigma}{2}+\mu^2\right) 
\pm
\frac{\mu\varsigma}{2}\exp\left(-\frac{\mu^2}{\varsigma}\right), \\	
\int_{-\infty}^\infty \chi_{\pm x>0} x^3\exp\left(-\frac{(x-\mu)^2}{\varsigma}\right)\diff x = \frac{\sqrt{\varsigma\pi}\mu}{2}\left(\frac{3\varsigma}{2}+\mu^2\right)\left(1\pm\mathrm{erf}\left(\frac{\mu}{\sqrt{\varsigma}}\right)\right)\\
\qquad\qquad\qquad\qquad\qquad\qquad\qquad\qquad
\pm
\frac{\varsigma}{2}\exp\left(-\frac{\mu^2}{\varsigma}\right)(\mu^2+\varsigma).
\end{gather*}
The first (un-centred) moment is then zero:
\begin{align}
&\int_{\mathbb{R}}\diff p_1 \mathcal{L}_{\mathrm{coll}}\left(f^{(1)},g^{(2)},t\right)=0.\label{eq:UncentredMom1}
\end{align}
We define
\begin{gather}
v_{\mathrm{diff}}^\pm = v(r,t)-v(r\pm\sigma,t),\quad
v_{\mathrm{sum}}^\pm = v(r,t)+v(r\pm\sigma,t),\\
E_{\mathrm{diff}}^\pm = E(r,t)-E(r\pm\sigma,t),\quad
E_{\mathrm{sum}}^\pm = E(r,t)+E(r\pm\sigma,t),\\
\rho^{\pm} = \rho(r\pm\sigma,t),\quad g_2^\pm=g_2(r\pm\sigma,t).	
\end{gather}
The second un-centred moment is then written in terms of error functions:
%
\begin{align}
\int_{\mathbb{R}}\diff p_1 p_1\mathcal{L}_{\mathrm{coll}}&\left(f^{(1)},g^{(2)},t\right)=
-\sum_{+,-}
\nonumber\frac{g_2^\pm\rho\rho^\pm\sqrt{mk_BTE_{\mathrm{sum}}^\pm}(1+\alpha)}{2\sqrt{2\pi}}
\exp\left(-\frac{m(v_{\mathrm{diff}}^\pm)^2}{2k_BTE_{\mathrm{sum}}^\pm}\right)
v_{\mathrm{diff}}^\pm
\\&
\pm\frac{g_2^\pm\rho\rho^\pm(1+\alpha)}{4}
\left(1-\mathrm{erf}\left(\frac{\sqrt{m}v_{\mathrm{diff}}^\pm}{\sqrt{2k_BTE_{\mathrm{sum}}^\pm}}\right)\right)
\left(k_BTE_{\mathrm{sum}}^\pm+m(v_{\mathrm{diff}}^\pm)^2\right)\label{eq:UncentredMom2}
\end{align}

And the third un-centred moment is given by

\begin{align}
&\int_{\mathbb{R}}\diff p_1 p_1^2\mathcal{L}_{\mathrm{coll}}\left(f^{(1)},g^{(2)},t\right)
=
-\sum_{+,-}
\frac{mg_2^\pm\rho\rho^\pm\sqrt{mk_BTE_{\mathrm{sum}}^\pm}}{2\sqrt{2\pi}}
\exp\left(-\frac{(mv_{\mathrm{diff}}^\pm)^2}{2k_BTE_{\mathrm{sum}}^\pm}\right)\nonumber
\\&\times\Bigg\{
(1+\alpha)\left(2k_BTE_{\mathrm{diff}}^\pm+mv_{\mathrm{diff}}^\pm v_{\mathrm{sum}}^\pm\right)
+\frac{1-\alpha^2}{2}\left(2k_BTE_{\mathrm{sum}}^\pm+m(v_{\mathrm{diff}}^\pm)^2\right)
\Bigg\}\nonumber\\&
\pm
\frac{m^2g_2^\pm\rho\rho^\pm}{4}
\left(1-\mathrm{erf}\left(\frac{\sqrt{m}v_{\mathrm{diff}}^\pm}{\sqrt{2k_BTE_{\mathrm{sum}}^\pm}}\right)\right)\nonumber
\\
&\times\Bigg\{
(1+\alpha)[v_{\mathrm{sum}}^\pm(k_BTE_{\mathrm{sum}}^\pm+m(v_{\mathrm{diff}}^\pm)^2)+2k_BTE_{\mathrm{diff}}^\pm v_{\mathrm{diff}}^\pm]
+\frac{1-\alpha^2}{2}v_{\mathrm{diff}}^\pm[3k_BTE_{\mathrm{sum}}^\pm+m(v_{\mathrm{diff}}^\pm)^2]
\Bigg\}\label{eq:UncentredMom3}
\end{align}
where the sum runs over $+$ and $-$ in place of $\pm$. The centred moments are then constructed as a linear combination of \cref{eq:UncentredMom1,eq:UncentredMom2,eq:UncentredMom3}.

\section{DDFT using an analytic approximation of $g$}
\begin{figure}
	\centering
	\includegraphics[width=\textwidth]{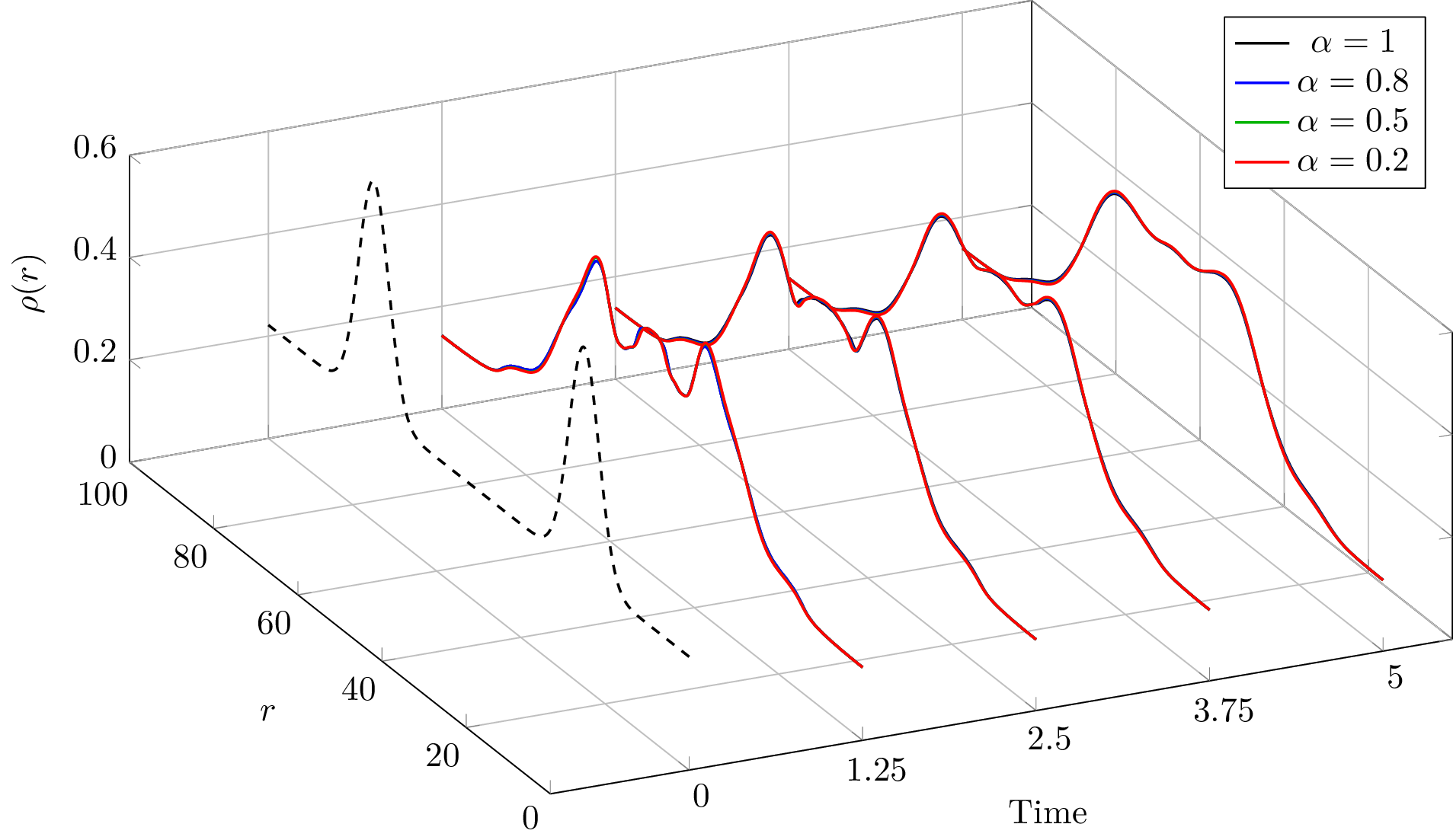}
	\caption{Results from the DDFT simulation, using $g^{(2)}(\sigma)$ given by \cref{eq:gRET}, using the same initial condition (black, dashed), for different coefficients of restitution $\alpha$.}\label{Figure:rhogRET}
\end{figure}

In \cite{Marini2007}, an analytic approximation of the radial correlation function is considered, which is independent of $\alpha$. Although our numeric investigation of the correlation function for different $\alpha$ disagrees with the analytic approximation, for comparison in \cref{Figure:rhogRET}, we provide results using this approximation of $g$, with the same initial configurations considered in \Cref{Section:Numerics}, in \cref{Figure:rhoDifferentAlpha}. The results show that, under this approximation, the introduction of inelasticity plays a much smaller role in the dynamics.  This is in contrast to the large effects seen in the microscopic simulations.

\end{document}